\documentclass[a4paper,twocolumn,11pt,accepted=2017-05-09]{quantumarticle}
\pdfoutput=1
\usepackage[utf8]{inputenc}
\usepackage[english]{babel}
\usepackage[T1]{fontenc}
\usepackage{amsmath}
\usepackage{amssymb}
\usepackage{braket}
\usepackage{hyperref}
\usepackage[backend=biber,sorting=none]{biblatex}
\usepackage{subcaption}
\usepackage{tikz}
\usepackage{lipsum}
\usepackage{float}

\makeatletter
\providecommand\firstofone[1]{#1}
\makeatother

\begin{document}

\title{Experimental demonstration of a coherent detector blinding attack on a real CV-QKD system}

\author{Daniel Pereira}
\affiliation{Optical Quantum Technologies, Center for Digital Safety \& Security,  AIT Austrian Institute of Technology GmbH, Giefinggasse 4, 1210 Vienna, Austria}
\email{daniel.pereira@ait.ac.at}
\orcid{0000-0001-8496-3063}
\author{Vana Pezelj}
\affiliation{Optical Quantum Technologies, Center for Digital Safety \& Security,  AIT Austrian Institute of Technology GmbH, Giefinggasse 4, 1210 Vienna, Austria}
\email{vana.pezelj@ait.ac.at}
\affiliation{ISAE-SUPAERO, avenue Marc Pélegrin 10, BP 54032 31055 Toulouse CEDEX 4 - France}
\affiliation{Faculty of Science, University of Split, Ruđera Boškovića 33, HR-21000 Split, Croatia}
\author{Florian Prawits}
\affiliation{Optical Quantum Technologies, Center for Digital Safety \& Security,  AIT Austrian Institute of Technology GmbH, Giefinggasse 4, 1210 Vienna, Austria}
\email{florian.prawits@ait.ac.at}
\orcid{0000-0001-5351-8178}
\author{Hannes Hübel}
\affiliation{Optical Quantum Technologies, Center for Digital Safety \& Security,  AIT Austrian Institute of Technology GmbH, Giefinggasse 4, 1210 Vienna, Austria}
\email{hannes.huebel@ait.ac.at}
\orcid{0000-0002-1173-1904}

\maketitle

\begin{abstract}
Continuous-variable quantum key distribution provides a theoretical unconditionally secure solution to distribute symmetric keys among users in a communication network.
However, the practical devices used to implement these systems are intrinsically imperfect, and, as a result, open the door to eavesdropper attacks.
In this work, we present a novel implementation of a coherent detector blinding attack, in which the eavesdropper hinders the capability of the receiver to properly estimate the channel parameters, hiding the impact of their collective attack.
Our results show that excess noise in excess of 2.5~SNU can be reliably hidden by the eavesdropper, thus demonstrating the feasibility of the attack.
We also discuss how our attack strategy can be further improved to allow for even stronger attacks (by using more advanced modulation formats), and propose some countermeasures to prevent it.
\end{abstract}

\section{Introduction}
Continuous Variables Quantum Key Distribution (CV-QKD) tackles the problem of the generation and distribution of symmetric cryptographic keys, without assuming any computational limitations on the adversary, while employing telecom compatible equipment~\cite{grosshans_continuous_2002, leverrier_theoretical_2009}.
For the purpose of security analysis, the observed excess channel noise is used to bound the information available to an eavesdropper, which demands a careful and precise estimation of this parameter~\cite{laudenbach_continuous-variable_2018}.
This estimation requires for the receiver to respond linearly to inputs from the channel~\cite{grosshans_continuous_2002}.
Recent developments show that CV-QKD receivers can be forced into a non-linear regime that would break this crucial assumption~\cite{qin_homodyne-detector-blinding_2018,Qin2016}
\par
QKD was first proposed in 1984, using the polarization of single photons as a coding basis, yielding the now well known BB84 protocol~\cite{bennett_quantum_2014}.
Nevertheless, the use of single photons poses difficulties in their practical implementation, namely the specialized equipment needed for single photon generation and detection~\cite{ralph_continuous_1999}.
Coherent-state CV-QKD typically encodes the information in the phase and amplitude of weak coherent states, thus allowing for implementation with current modulation methods and telecom-based equipment and techniques~\cite{grosshans_continuous_2002,almeida_secret_2021}.
One such telecom-based technique is the usage of balanced (also called coherent) detectors, which allows for the recovery of the phase information encoded in the signal~\cite{loudon_quantum_2000}.
The security of QKD systems is usually proven under the assumption of ideal devices and components, with little to no consideration to the actual implementation~\cite{scarani_security_2009}.
Early on, this assumption was challenged with the proposal of the Photon Number Splitting (PNS) attack, which exploits the fact that realistic DV-QKD systems employ Weak Coherent States (WCS).
WCS's photon number follows a Poisson distribution~\cite{loudon_quantum_2000}, which means they sometimes contain more than one photon, allowing for an eavesdropper to extract those surplus photons to gain information without being detected~\cite{huttner_quantum_1995}.
Countermeasures to this attack were proposed, such as the use of decoy states, in which additional states with different mean photon numbers (resulting in different Poissonian distributions) are introduced to the signal at random, undisclosed points in time.
These two facts (the existence of multiple Poissonian distributions and the secrecy, while in transit, to which distribution any individual state belongs to), means that the eavesdropper can no longer perfectly reconstruct the expected statistics of all distributions while introducing a photon-number dependent loss, such as is the case in the PNS attack, with the attack being detected through monitoring deviations in those detection statistics~\cite{hwang_quantum_2003, lo_decoy_2005}.
Many other attacks have been proposed since then, either exploiting the imperfections of devices used in the implementation or by manifesting them through fault injection~\cite{bsi_ais_2024}.
Early side-channel attacks in CV-QKD stemmed from the fact that, in the originally proposed CV-QKD systems, the Local Oscillator (LO) was transmitted from the transmitter to the receiver through the same channel as the signal, allowing for an eavesdropper to manipulate it (for example changing its pulse shape or wavelength) and thus hide their interference with the signal itself~\cite{jouguet_preventing_2013, Qi2015,ma_wavelength_2013,huang_quantum_2013, huang14}.
Again, countermeasures to these attacks were suggested, one such countermeasure being generating the LO locally at the receiver~\cite{Qi2015,soh15}, nowadays Local LO (LLO) systems are the norm.
Studies have been done on the impact of receiver imbalances on the security of CV-QKD, where it was shown that improperly balanced receivers will report wrong estimates of the secret key rate values~\cite{pereira_impact_2021} due to their inability of accurately measuring the channel parameters.
Further, it has been shown that coherent receivers can be forced into a saturated state by an eavesdropper, either by displacing the states from Alice~\cite{Qin2016, qin_saturation_2013} or by injecting a strong incoherent light into the receiver~\cite{qin_homodyne-detector-blinding_2018}.
These processes are interchangeably called saturation attacks or blinding attacks, and they have the potential to break the security of CV-QKD by hindering the receiver's ability to properly estimate the channel parameters, thus hiding the impact of an eavesdropper's attack on these parameters.
However, a detector blinding attack has not been launched against a full CV-QKD receiver.
\par
In this paper we present a novel implementation of a coherent detector blinding attack against a complete CV-QKD receiver.
Additionally, we describe a controlled wideband noise source to emulate the impact of a collective attack on the signal, and a blinding source to hide that noise.
We show that our noise source can reliably introduce a controlled amount of noise into the system in a reproducible way, and that our blinding source can work even against receivers with AC-coupling.
We show that the channel excess noise can be completely hidden by the blinding source, even for  excess noise values of $\approx$ 2.5~SNU.
\par
This paper is structured into 8 sections, including this introduction.
In \autoref{sec:practical_security} we present a quick note on how security is estimated in CV-QKD, identifying the assumptions that are relevant for the attack being implemented.
In \autoref{sec:attack_strategy} we present the general strategy of the attack being implemented, as well as specifying the requirements for the blinding stage of the attack.
In \autoref{sec:DUT} we present the details of the CV-QKD receiver being used as a target for the attack.
Afterwards, in \autoref{sec:noise_source} we present the details of the noise source that was assembled to simulate the impact of a collective attack on the channel parameters, and in \autoref{sec:blinding_source} we present the details of the blinding source that was assembled to achieve the enabling stage of the attack.
In \autoref{sec:attack_results} we present the results of the attack, showing that the excess noise introduced by the noise source can be hidden by the blinding source, and thus demonstrating the feasibility of the attack.
Finally, in \autoref{sec:conclusion} we summarize our findings, propose some countermeasures and present some concluding remarks.

\section{Practical security of CV-QKD}\label{sec:practical_security}
The security of CV-QKD against collective attacks was established in~\cite{leverrier_theoretical_2009}, for both Gaussian Modulated (GM) and Discrete Modulated (DM) protocols, with the DM proof having been extended for arbitrary constellation formats {in~\cite{denys_explicit_2021}}.
The achievable secure key rate is given by~\cite{leverrier_theoretical_2009}
\begin{equation}\label{eq:keyRate}
K=\beta I_\text{BA}-\chi_\text{BE},
\end{equation}
where $\beta$ is the reconciliation efficiency, $I_\text{BA}$ is the mutual information between Bob and Alice, given by~\cite{leverrier_theoretical_2009}:
\begin{equation}
I_\text{BA} = \log_2\left(1+\frac{2T\eta\braket{n}}{2+T\eta\epsilon+2\epsilon_\text{th}}\right),
\end{equation}
where $T$ is the channel transmission, $\eta$ is the efficiency of the receiver, $\braket{n}$ is the average number of photons per symbol, $\epsilon$ is the excess channel noise and $\epsilon_\text{th}$ is the thermal noise of the receiver.
In~\autoref{eq:keyRate}, $\chi_\text{BE}$ describes the Holevo bound that majors the amount of information that Eve can gain on Bob's recovered states, being obtained from the symplectic eigenvalues of the system's covariance matrix~\cite{leverrier_theoretical_2009}:
\begin{equation}\label{eq:covMat}
\gamma_\text{AB}=
\begin{bmatrix}
V\mathbb{I}_2 & \sqrt{T}Z\sigma_Z \\
\sqrt{T}Z\sigma_Z & (TV+ 1-T + T\epsilon)\mathbb{I}_2
\end{bmatrix},
\end{equation}
where $V = 2\braket{n} +1$ corresponds to the variance of the signal at the output of the transmitter plus the unavoidable shot noise, $\mathbb{I}_2$ is the 2-D identity matrix, ${\sigma_Z=\text{diag}(1,-1)}$, and where $Z$ is a measure of the correlation between the states at the transmitter and receiver, being given by~\cite{denys_explicit_2021}
\begin{equation}\label{eq:Zdefinition}
Z = 2\text{tr}(\hat\rho^{\frac{1}{2}}\hat{a}\hat\rho^{\frac{1}{2}}\hat{a}^\dagger)-\sqrt{2\epsilon W},
\end{equation}
where $\hat{\rho}=\sum_{k=1}^M p_k\ket{\alpha_k}\bra{\alpha_k}$ is the density operator of the M-symbol discrete constellation and~\cite{denys_explicit_2021}
\begin{equation}
W = \sum_{k=1}^Mp_k(\bra{\alpha_k}\hat{a}^\dagger_\rho\hat{a}_\rho\ket{\alpha_k}-|\bra{\alpha_k}\hat{a}_\rho\ket{\alpha_k}|^2),
\end{equation}
and finally $\hat{a}_\rho=\hat\rho^{\frac{1}{2}}\hat{a}\hat\rho^{-\frac{1}{2}}$.
The exact methodology to compute $\chi_\text{BE}$ can be found {in~\cite{denys_explicit_2021}}.
\par
To estimate the channel parameters (transmission, $T$, and excess noise, $\epsilon$), security proofs consider that  Alice and Bob share a couple of correlated variables $a$ and $b$, corresponding to the states generated by Alice and those received by Bob, respectively.
Both variables are assumed to be expressed in Shot Noise Units (SNU), for $a$ this is accomplished by setting the constellation amplitude so that the expected value of $a$, $E[a]$, is equal to $\sqrt{2\braket{n}}$, while for $b$ it is done by having Bob divide the output of his Digital Signal Processing (DSP) stage by the shot noise variance.
$a$ and $b$ are related by the normal linear model~\cite{leverrier_theoretical_2009}:
\begin{equation}\label{eq:linearModel}
b=ta+z,
\end{equation}
where ${t=\sqrt{\eta T}}$ and $z$ is the noise contribution, following a normal distribution with null mean and variance 
\begin{equation}\label{eq:variance}
\sigma^2=1+\epsilon_\text{th}+\eta T\epsilon
\end{equation}
$t$ and $\sigma^2$ can be estimated through~\cite{kleis_continuous_2017}:
\begin{equation}
\tilde{t} = \text{Re}\left\lbrace\frac{\sum_{i=1}^Na_ib_i^*}{N}\right\rbrace,\qquad
\tilde{\sigma}^2=\frac{\sum_{i=1}^N|b_i-\tilde{t}a_i|^2}{N}.
\end{equation}
The transmission and excess noise are then estimated through:
\begin{equation}
\tilde{T} = \frac{\tilde{t}^2}{2\braket{n}\eta},\qquad\tilde{\epsilon}=\frac{\tilde{\sigma}^2-1-\epsilon_\text{th}}{\eta \tilde{T}},
\end{equation}
where the $\tilde{}$ notation is used to denote that these are now estimates of the actual values.
For the results presented in this work, no finite size effects are taken into consideration.
\par
The linear model in \autoref{eq:linearModel} is only valid when the coherent detector is operating in its linear regime, i.e., when the output voltage is proportional to the input optical power.
If the detector is blinded, this linear relationship is no longer valid, and the estimates $\tilde{t}$ and $\tilde{\sigma}^2$, and consequently $\tilde{T}$ and $\tilde{\epsilon}$, will be incorrect.

\section{Coherent Detector Blinding Attack Strategy}\label{sec:attack_strategy}
\begin{figure}[h]
\includegraphics[width=\linewidth]{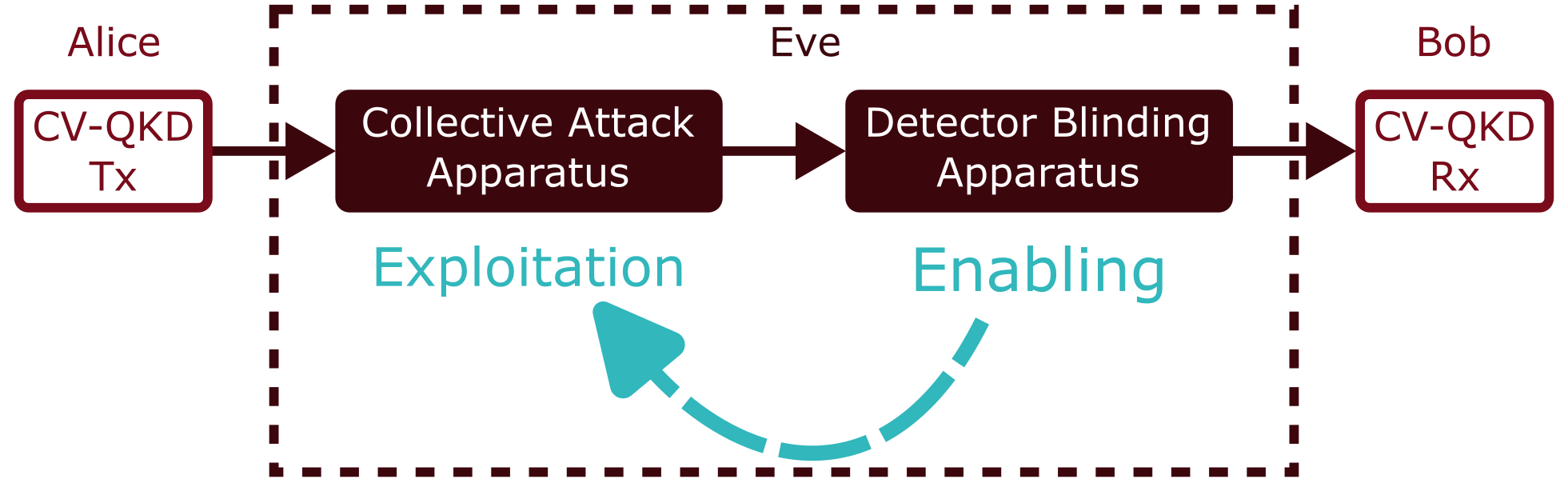}
\caption{Diagram of the general attack strategy being considered in this paper.}
\label{fig:attack_strategy}
\end{figure}

A simplified schematic of the attack scenario as would be implemented by an adversary is presented here in \autoref{fig:attack_strategy}, where the two different stages of the attack are clearly labelled.
The attack consists of two stages, the enabling stage, in which the adversary blinds the detector, and the exploitation stage, in which the adversary performs a collective attack, taking advantage of the fact that the impact of their tampering on the channel parameters is hidden by the blinding.
\par
For the purposes of this paper, we assume that the exploitation stage implemented by the adversary consists of a collective attack, a strategy which has been shown to be optimal in the asymptotic limit~\cite{leverrier_theoretical_2009}.
The details of the exploitation stage are not relevant for the purposes of this paper, as we are only interested in demonstrating the feasibility of the enabling stage, and showing that it can be used to hide the impact of an attack on the channel parameters.
\par
The enabling stage consists of the adversary sending a strong optical signal to the receiver, with the goal of forcing the detector into a non-linear regime, in which the output voltage is no longer proportional to the input optical power.
Of particular importance is the choice of the wavelength of the blinding signal, for two reasons:
\begin{enumerate}
    \item The adversary needs to be able to achieve a differential illumination of the balanced detector, essentially sending considerably more power to one of the photodiodes than to the other, which causes the output to swing into one of the limits (upper or lower) of the receiver response.
    A simplified schematic diagram of this is presented in \autoref{fig:Blinding_Schematic}.
    \begin{figure}[h]
        \centering
        \includegraphics[width=\linewidth]{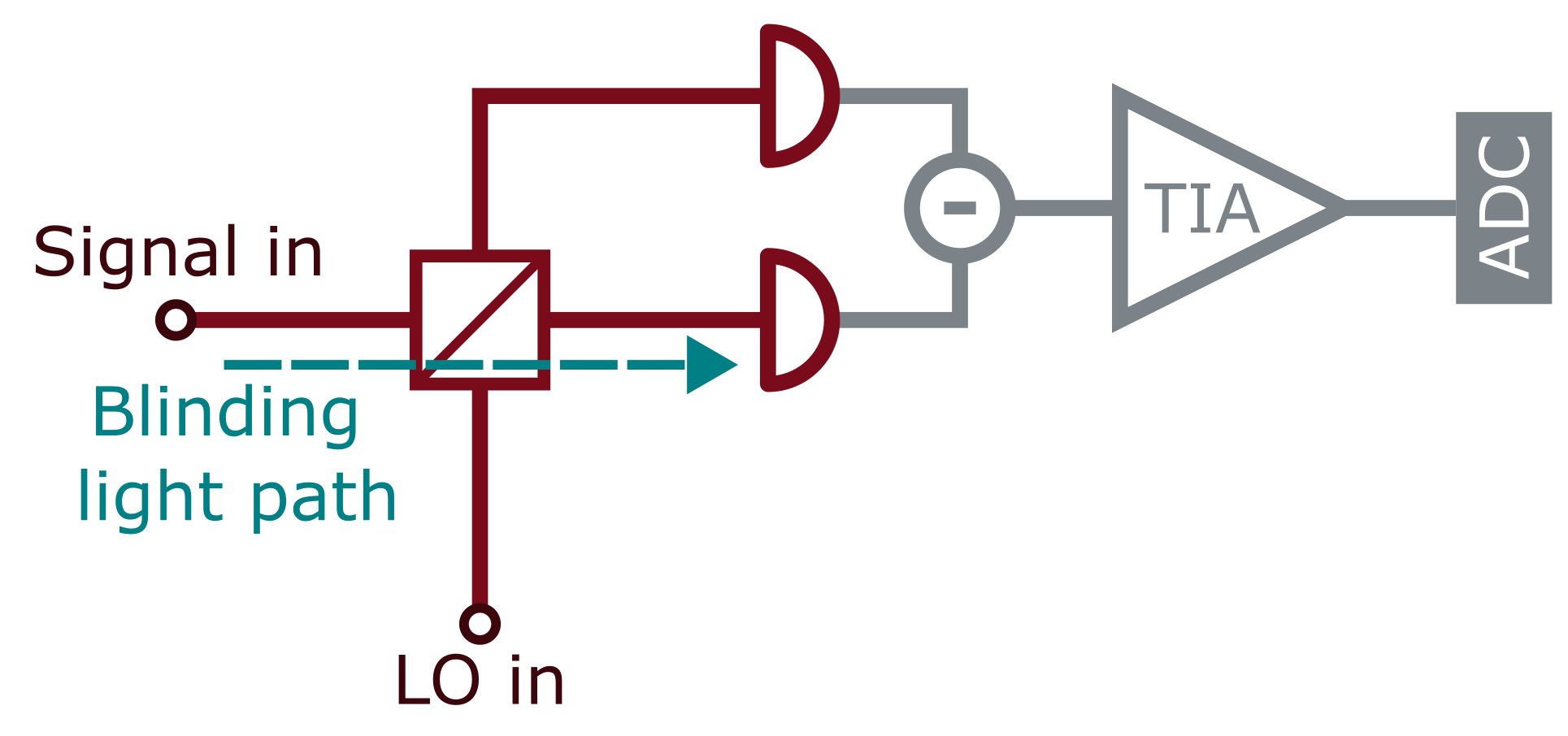}
        \caption{Schematic diagram of the required path for the blinding light.}
        \label{fig:Blinding_Schematic}
    \end{figure}
    This can be achieved by exploiting the wavelength dependence of the components inside the receiver being attacked, specifically of the 50/50 BS.
    \item The interference signal between the blinding signal and the LO needs to be negligible, otherwise this signal will interfere with the normal operation of the receiver (for example, by causing the DSP to lose lock on the pilot), turning the attack into a denial of service attack, rather than a stealthy attack on the channel parameters.
    This can be achieved by detuning the blinding signal from the LO by enough to ensure that the interference signal is outside the electronic bandwidth of the balanced detector, and thus cannot be detected by the receiver.
\end{enumerate}
Around 1550~nm, a detuning of 1~nm is equivalent to a frequency shift of approximately 125~GHz, a value that is far beyond the bandwidth of common balanced detectors, which is usually in the low GHz range~\cite{paschotta_balanced_2019}.
Meanwhile, the wavelength stability range of a BS is usually in the order of a few tens of nanometers, with large imbalances only being observed at wavelengths $>100~\text{nm}$ distant from the operating wavelength that the BS is designed for~\cite{digonnet_wavelength_1983}.
Additionally, the photodiodes of the balanced detectors usually have an operating wavelength range of a few hundreds of nanometers, outside which the responsivity drops significantly, hindering the achievement of saturation.
As a combination of these factors, the blinding signal should be detuned from the LO by a value that is large enough to ensure that a significant imbalance in the BS is observed but while retaining a wavelength value that is still within the operating range of the photodiodes.
\par
In operational terms, blinding a detector means deviating its output by such a degree that it reaches some maximum value.
This maximum value can be determined by multiple components, such as the saturation of the photodiodes, the saturation of the Transimpedance Amplifier (TIA) or the clipping of the output voltage by the Analog to Digital Converter (ADC).
In mathematical terms, given an input voltage \(V\) and a receiver with upper and lower limits \(V_\text{UL}\) and \(V_\text{LL}\), this can be expressed as
\begin{equation}\label{eq:saturation}
    V_\text{out} = 
    \begin{cases}
        V_\text{UL}, &V\geq V_\text{UL} \\
        V, & V_\text{LL} < V < V_\text{UL} \\
        V_\text{LL}, & V \leq V_\text{LL}
    \end{cases}.
\end{equation}
Essentially, \autoref{eq:saturation} describes how the receiver becomes non-injective, i.e. due to the limiting behaviour in the output range of some part of the receiver, the linear relationship of inputs and outputs breaks down and distinct inputs can be maxed to the same output value (lower or upper limit).
\par
This effect can be further exacerbated by the fact that close to this saturation point, the output voltage can be highly non-linear.
This means that even if the detector is not fully blinded, the linear model used for the estimation of the channel parameters is still not valid, and thus the estimates will be incorrect.

\begin{figure*}[!t]
    \centering
    \includegraphics[width=0.80\linewidth]{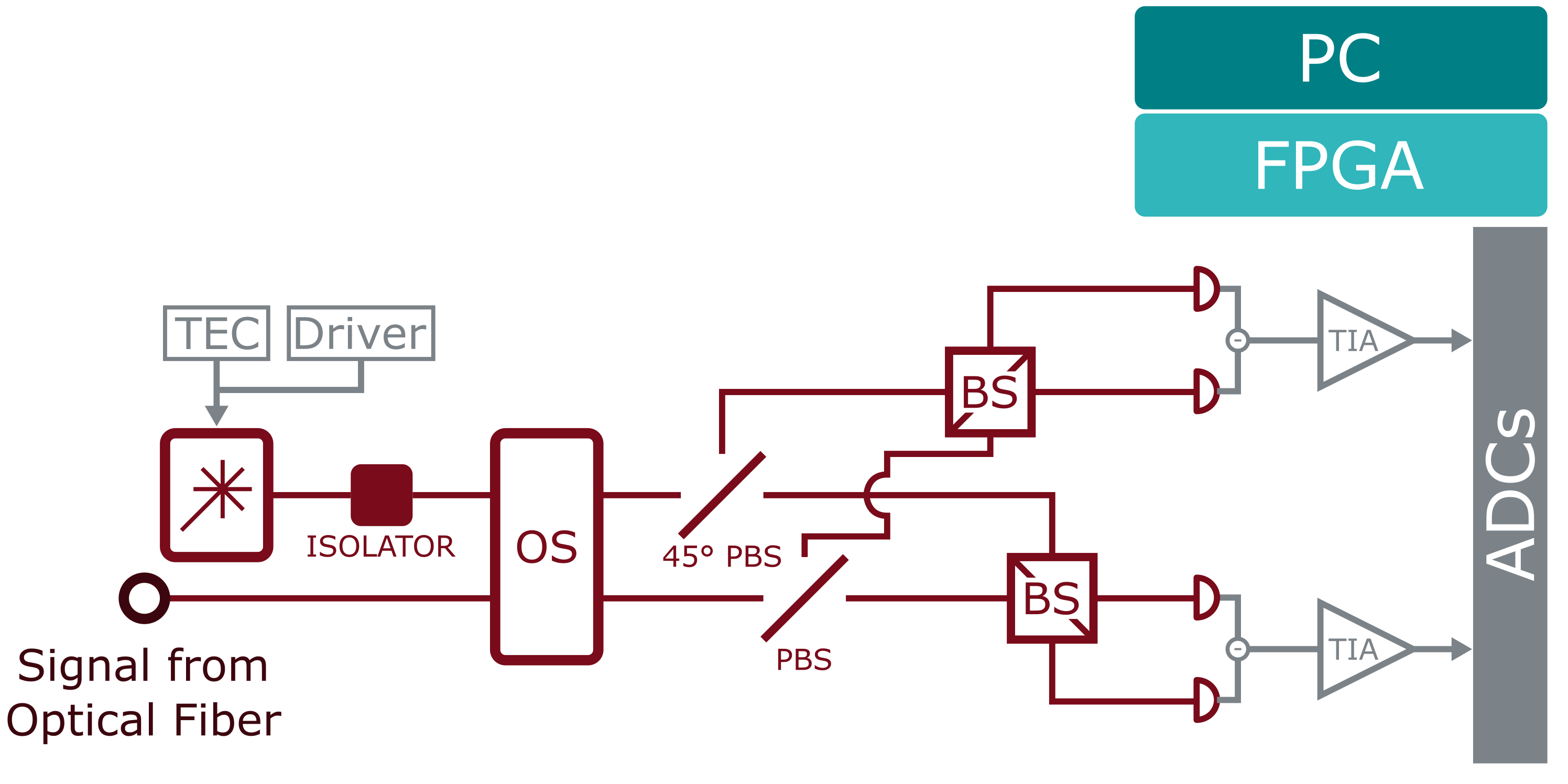}
    \caption{Diagram of the CV-QKD receiver being used as a target for the attack. ADC: Analog to Digital Converter; BS: 50/50 Beam Splitter; FPGA: Field Programmable Gate Array; OS: Optical Switch; PBS: Polarizing Beam Splitter; TEC: TEmperature Controller; TIA: Trans-Impedance Amplifier.}
    \label{fig:DUT_diagram}
\end{figure*}

\section{Description of system under attack}\label{sec:DUT}

\begin{figure}[h]
    \centering
    \includegraphics[width=.8\linewidth]{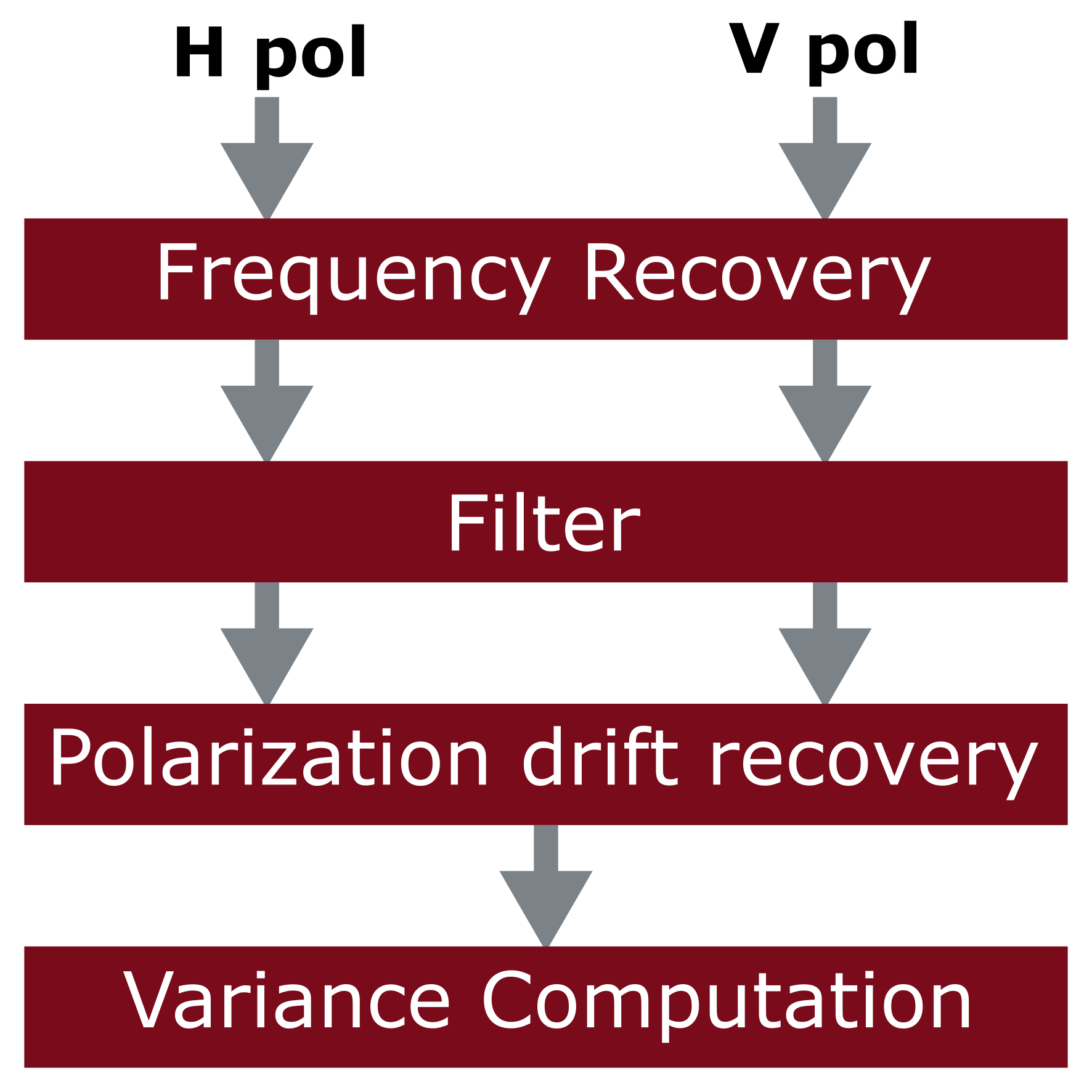}
    \caption{Diagram of the DSP stage of the CV-QKD receiver.}
    \label{fig:DSP}
\end{figure}

The attack strategy illustrated in \autoref{fig:attack_strategy} incorporates a full CV-QKD system, however for the portion of the attack being implemented here, the enabling/blinding stage, only the receiver (Bob) is relevant, as the attack is launched against it without any tampering of the transmitter (Alice).
Because of this, only the receiver was considered for this publication.
A diagram of the CV-QKD receiver being targeted is shown in \autoref{fig:DUT_diagram}. 
The receiver laser, taking the role of the local oscillator (LO), consists of an \textit{Aerodiode 1550nm DFB Model 5} model, being run by a \textit{Wavelength Electronics PTC2.5K-CH} temperature controller and a \textit{Wavelength Electronics MPL250} laser driver.
The optical output of this laser is fed through an optical isolator, followed by an \textit{Agiltron LightBend 1x1 PM} optical switch, and then into the polarization diverse receiver assembly, where it will be mixed with the signal from the fibre. 
The signal incoming from the fibre is first sent through an \textit{Agiltron LightBend 1x1} optical switch, and then into the polarization diverse receiver assembly. 
The two optical switches are used to allow for shutting off the signal and LO paths independently, allowing for on the fly estimation of both thermal noise (when both paths are shut off) and shot noise (when only the signal path is shut off).
In the polarization diverse receiver, the receiver laser signal (the LO) is first passed a PBS, with its fast-axis shifted 45° in relation to the polarization alignment of the laser, effectively sending half the power to each individual 50/50 BS. 
The quantum signal is also passed through its own PBS, splitting the two polarizations components, with each being sent to the different 50/50 BS. 
Both 50/50 BSs are polarization maintaining, ensuring that the polarization of both the signal and LO  remain aligned in each.
The outputs of each 50/50 beam-splitter are fed into a pair of \textit{Wieserlabs WL-BPD1GA} balanced optical receivers, that have a bandwidth of 1.6~GHz and DC cut-off of frequencies below 300~kHz~\cite{noauthor_wl-bpd1ga_2025}.
This doubled balanced receiver configuration allows for the measurement of both polarization components of the incoming signal, for the purposes of polarization drift compensation that is applied digitally.
The outputs of the two balanced receivers are then digitized by a \textit{Texas Instruments ADC12DJ3200EVM} ADC, running at a sample rate of 3.2~GHz.
The digitized signals are then sent to a PC, where they are subjected to the Digital Signal Processing (DSP) stage presented, in diagram form, in \autoref{fig:DSP}.
For the purposes of this paper, a reduced version of the DSP stage is used.
The two signals are identified as \textbf{H pol} and \textbf{V pol} in \autoref{fig:DSP}, as they correspond to the horizontal and vertical polarization components.
The DSP runs as follows:
\begin{enumerate}
    \item Frequency recovery is performed, downconverting the area of the signal around 400~MHz to baseband (accomplished by multiplying the time-domain signal by $e^{-i2\pi 400\times10^6 t}$). This is done for both polarization components independently.
    \item The frequency recovered signal is then filtered by a Root Raised Cosine (RRC) filter with a roll-off factor of 0.95 and a symbol rate of 153.6~MBaud, this step follows the same logic as would be used in the full DSP. This is done for both polarization components independently.
    \item Polarization drift recovery is then performed by combining the two polarization components of the signal, assuming the polarization components are equally distributed between the two arms of the receiver, using the method described in~\cite{pan_simple_2023}.
    \item Finally, the variance of the polarization recovered signal is computed.
\end{enumerate}
This reduced version of the DSP differs from the full version, which can be found in \cite{pereira_polarization_2023}, in a few ways, mainly by hard coding some values that would be obtained from the pilot tone in the full version (such as the downconversion frequency and the polarization combination coefficients~\cite{pereira_polarization_2023}), and by skipping some steps entirely (such as the phase compensation, clock recovery and frame reconciliation steps).
The reduction is done both to simplify the implementation and because, due to not using the transmitter in the experiment, neither the pilot tone nor the signal are present, making those steps either unnecessary or impossible to perform.
\par
Using the optical switches, \textit{thermal} (when both switches are closed), \textit{shot} (when only the LO path is open) and \textit{noisy} (when both switches are open) signals are acquired and subjected to the DSP, their variances being returned as \text{Var[thermal]}, \text{Var[shot]} and \text{Var[noisy]}, respectively.
The parameters in \autoref{eq:variance} can then be obtained from these variances.
The shot noise variance is obtained by subtracting the thermal noise variance from the variance of the shot signal (note that when acquiring the shot noise signal, the thermal noise is still present), $\text{SNU}=\text{Var[shot]}-\text{Var[thermal]}$.
The two noises constitute two independent random variables, thus to obtain the variance of the shot noise a subtraction of the thermal noise variance is sufficient.
The variance of the $z$ parameter in \autoref{eq:linearModel}, $\tilde{\sigma}^2$, is obtained by normalizing the variance of the noisy signal by the SNU:
\begin{equation}
    \tilde{\sigma}^2 = \frac{\text{Var[noisy]}}{\text{SNU}}.
\end{equation}
The thermal noise variance, $\epsilon_\text{th}$, is obtained by normalizing the variance of the thermal signal by the SNU:
\begin{equation}
    \epsilon_\text{th} = \frac{\text{Var[thermal]}}{\text{SNU}}.
\end{equation}
The lack of a transmitter in the experiment means that the channel transmission, $T$, cannot be estimated (the values of $2$ in \autoref{eq:linearModel} are essentially all 0), for the purposes of this paper we assume a transmission of $T=0.5$, equivalent to 15km of standard single mode fibre.
$\eta$ can be obtained from the specifications of the balanced receiver, being $\eta=0.90$ for the \textit{Wieserlabs WL-BPD1GA} model~\cite{noauthor_wl-bpd1ga_2025}.
Plugging all these values into \autoref{eq:variance} yields an estimate of the excess noise $\tilde{\epsilon}$.

\section{Noise Source}\label{sec:noise_source}

\begin{figure}[h]
    \centering
    \includegraphics[width=.9\linewidth]{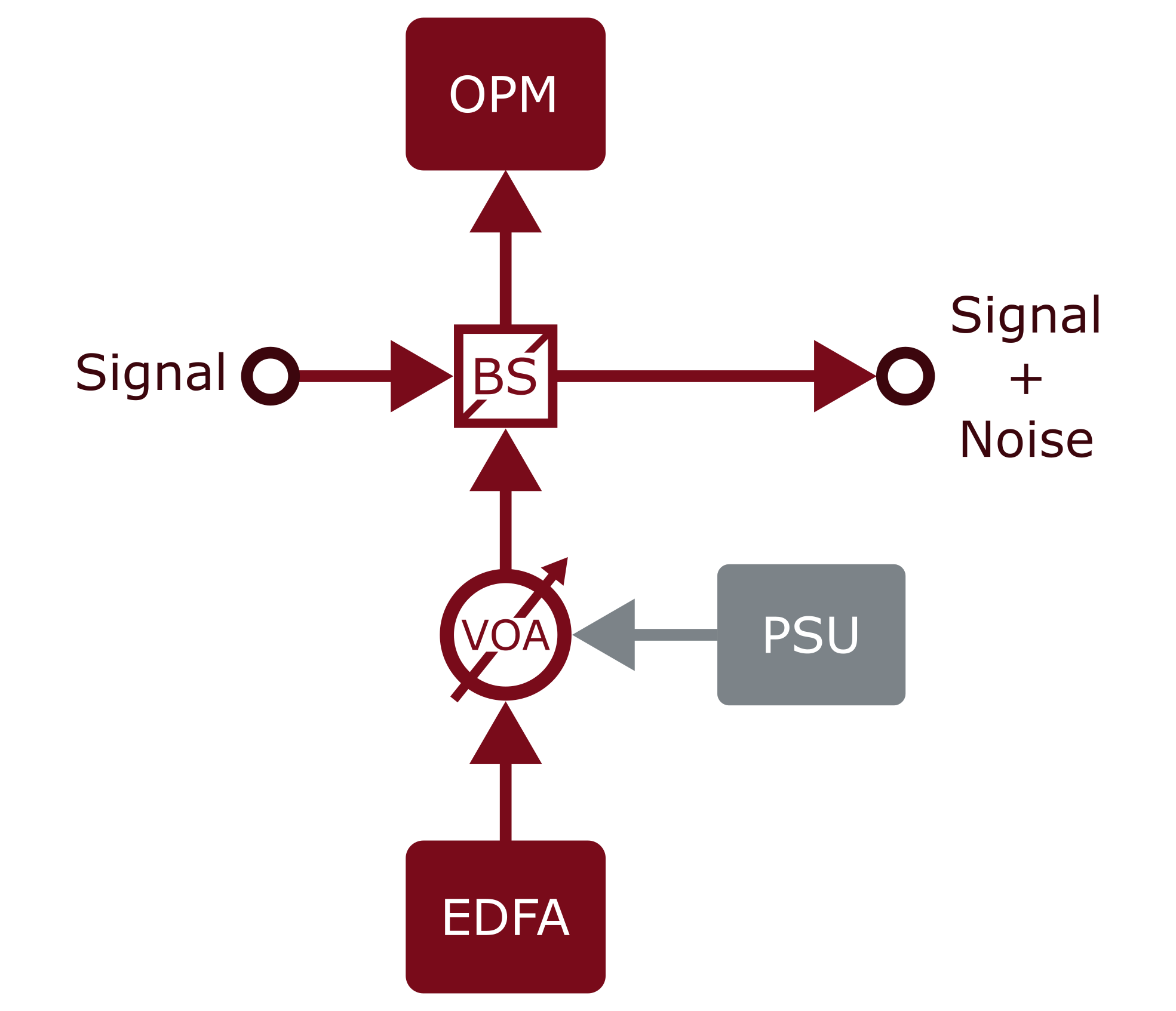}
    \caption{Diagram of the noise source used to simulate the impact of the collective attack. BS: 90/10 Beam Splitter; EDFA: Erbium-Doped Fibre Amplifier; OPM: Optical Power Meter; PSU: Power Supply Unit; VOA: Variable Optical Attenuator.}
    \label{fig:noise_source}
\end{figure}

To simulate the presence of an eavesdropper performing a collective attack, we assembled a noise source that allows us to introduce a controlled amount of incoherent noise into the system.
A diagram of this noise source is shown in \autoref{fig:noise_source}.
The incoherent signal is obtained from a \textit{ThorLabs EDFA100S} Erbium-Doped Fibre Amplifier (EDFA)  with no input signal, thus exploiting its Amplified Spontaneous Emission (ASE).
The output of the EDFA is then passed through a \textit{ThorLabs V1550A}  voltage controlled Variable Optical Attenuator (VOA), with the control voltage being supplied by a UNI-T \textit{UDP3305S} Power Supply Unit (PSU), which in turn was controlled by the PC via serial port.
After attenuation, the noise signal is then combined with the signal from the transmitter by the use of a \textit{WBC-2x2-13/15-10/90} 90/10 Beam Splitter (BS), using a port combination that redirected most of the noise signal from the EDFA to an OZ-Optics \textit{POM-300} Optical Power Meter (OPM).
\par
\begin{figure}[h]
    \centering
    \includegraphics[width=\linewidth]{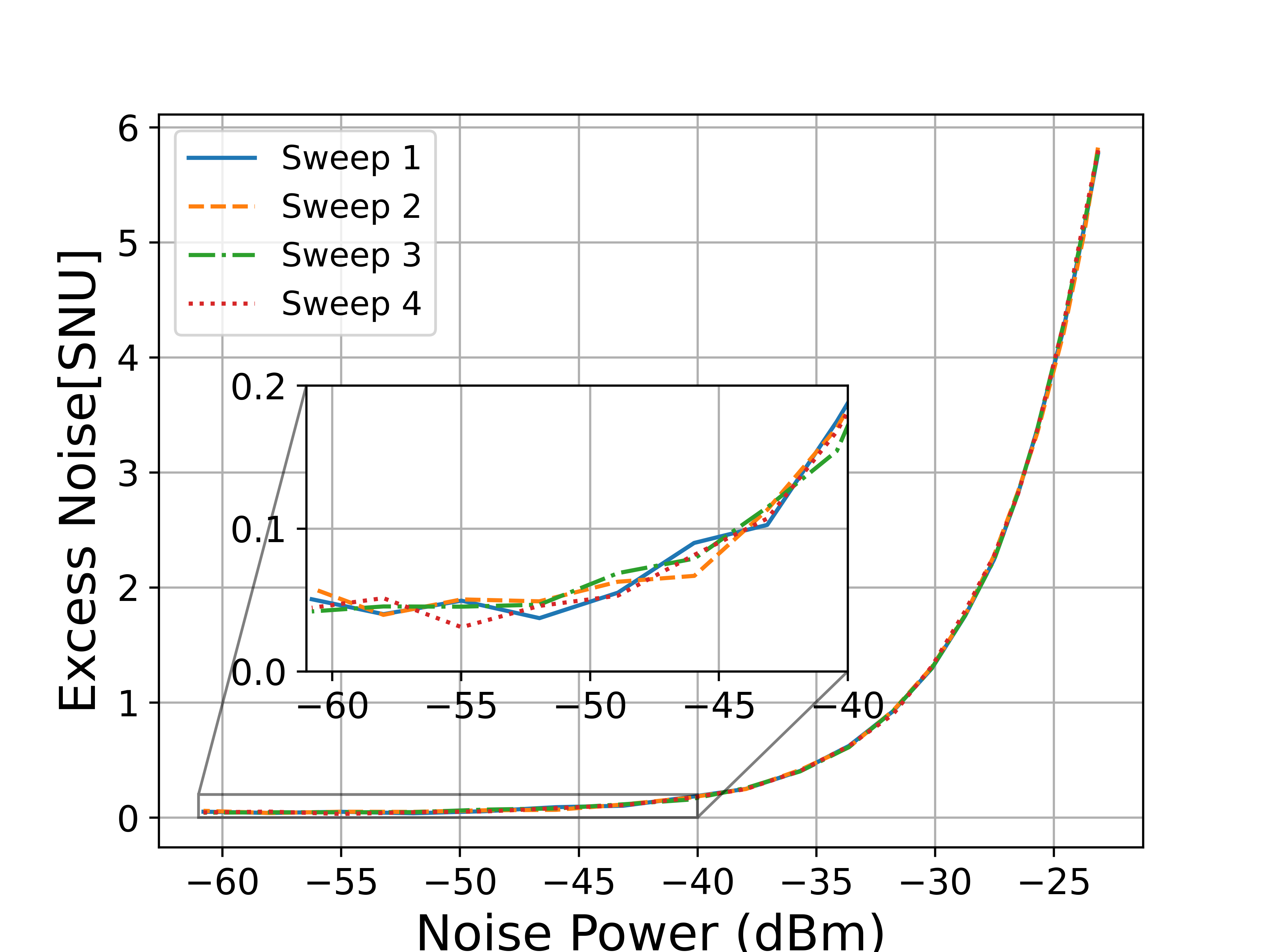}
    \caption{Demonstration of the reproducibility of the noise source.}
    \label{fig:noise_source_demonstration}
\end{figure}

To test the reproducibility of the noise source, we acquired 4 power sweeps, in which the power of the noise signal (controlled by the VOA) was varied from -13.6~dBm to -51.25~dBm in steps of varying size (narrower steps for higher attenuation).
The OPM readings were taken at the 90\% output of the BS and their value was then converted to the value of the 10\% using the known splitting ratio.
The results, in the form of the excess noise reported by the receiver and estimated following the method explained at the end of \autoref{sec:DUT}, of this evaluation are shown in \autoref{fig:noise_source_demonstration}.
The results show that the noise source is highly reproducible, with the curves from all 4 sweeps overlapping.
The inset in \autoref{fig:noise_source_demonstration} shows a zoomed in view of the region between -61~dBm and -40~dBm, where the different sweeps are now discernible, but it is still clear that the noise source maintains a high level of consistency.
\par
The EDFA has a bandwidth of roughly 35~nm, centered on 1550~nm.
Given that the OPM was programmed to report power values for a 1550~nm wavelength, the values it returns won't be extremely accurate.
Since the OPM is only used to aid in the control of the noise source (the parameter that matters is the excess noise reported by the receiver), this doesn't pose a problem.


\section{Blinding Source}\label{sec:blinding_source}
\begin{figure}[h]
    \centering
    \includegraphics[width=.9\linewidth]{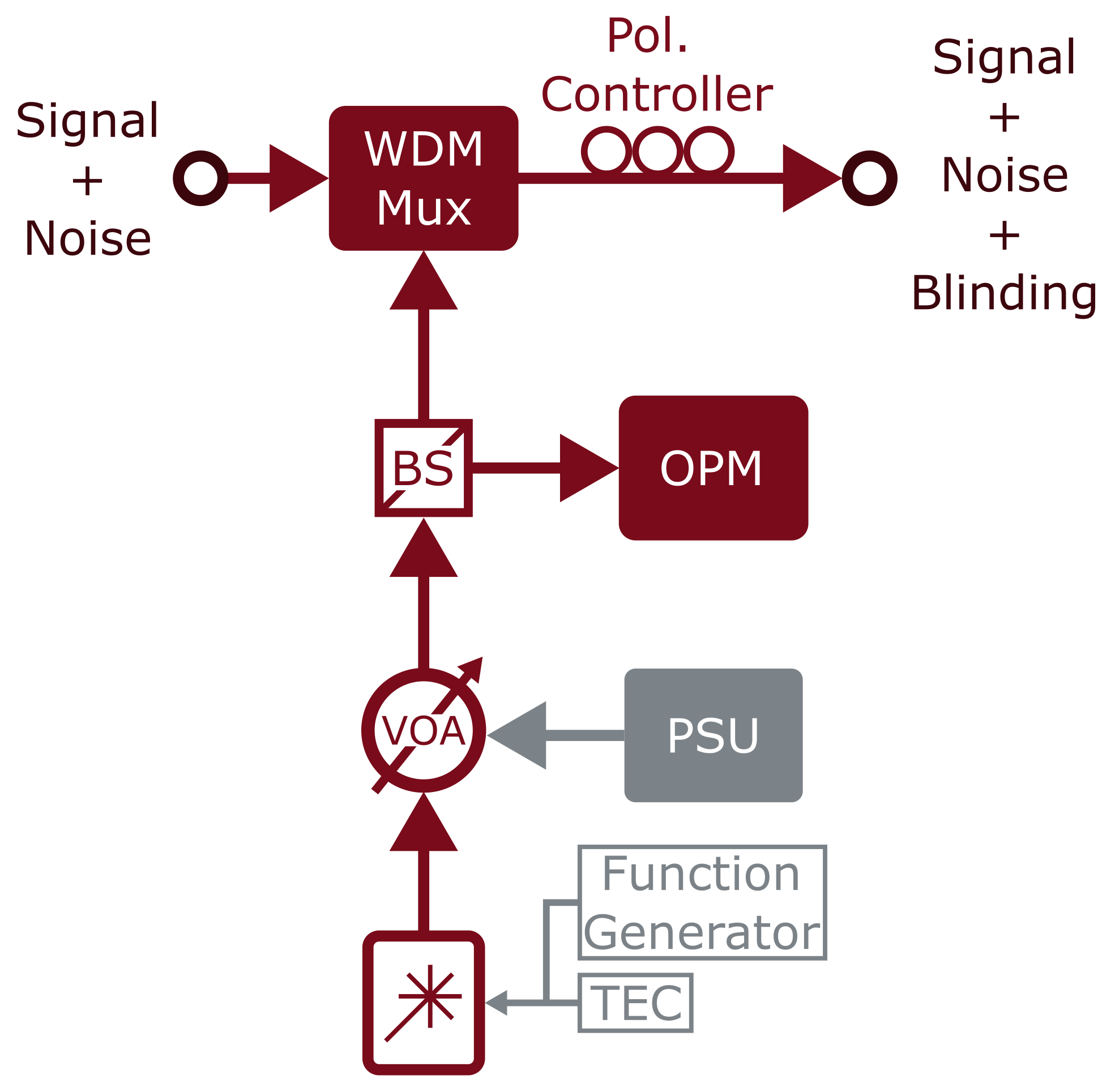}
    \caption{Diagram of the blinding source stage of the attack. BS: 50/50 Beam Splitter; OPM: Optical Power Meter; PSU: Power Supply Unit; TEC: TEmperature Controller; WDM: Wavelength Division Multiplexer; VOA: Variable Optical Attenuator.}
    \label{fig:blinding_source}
\end{figure}

To achieve the blinding of the detector, the blinding source presented in \autoref{fig:blinding_source} was assembled.
The source itself consisted of a \textit{Sumitomo Electric Industries SLV5410} diode emitting at 1344~nm, being run by a \textit{Rigol DG1032} function generator, which was used to supply the current to the laser diode, and a \textit{Thorlabs TED200C} temperature controller.
The function generator was used to achieve direct modulation of the laser, which was necessary to bypass the AC-coupling of the balanced detector, which would otherwise filter out any constant blinding signal, and thus prevent the achievement of saturation.
A square modulation pattern at a frequency of 10~MHz was used, well above the 300~kHz cut-off frequency of the balanced detector.
The modulated blinding signal was then sent through a \textit{ThorLabs V1550A} VOA, with the control voltage being supplied by a UNI-T \textit{UDP3305S} PSU, which in turn was controlled by the PC via serial port.
The power controlled signal was then sent through a \textit{Thorlabs TD1315R5F1} 50/50 BS (dual window 1310/1550~nm), with one output being sent to an OZ Optics \textit{POM-600} OPM, and the other being sent to a \textit{Thorlabs WD1350B} Wavelength Division Multiplexer (WDM).
In this WDM the blinding signal was combined with the signal in the fibre, with the combined signal then being passed through a polarization controller.
The polarization controller was used to fine tune the polarization of the blinding signal, with the objective of providing the same amount of blinding power to both arms of the receiver.
This fine-tuning was accomplished by changing the polarization of the blinding signal while monitoring the output of the balanced detectors, with the optimal polarization being the one that resulted in the same amount signal amplitude output of both balanced receivers.
\par
\begin{figure}[h]
    \centering
    \includegraphics[width=\linewidth]{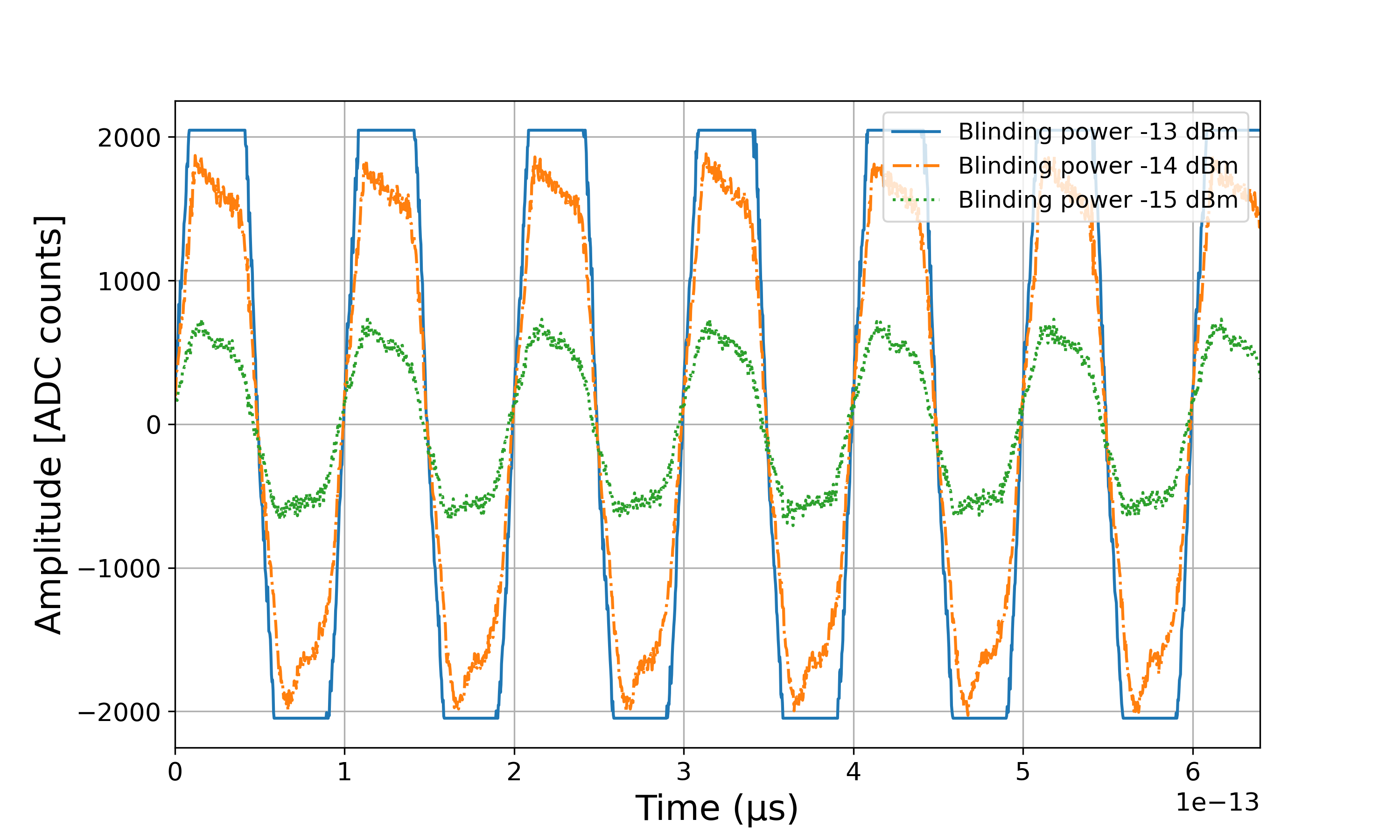}
    \caption{Impact of increasing blinding power on the linearity of the Horizontal receiver.}
    \label{fig:blinding_demonstration}
\end{figure}

To illustrate the action of the blinding source, below in \autoref{fig:blinding_demonstration} we show the output of the balanced detector in horizontal arm of the receiver as the power of the blinding signal is increased from -15~dBm to -13~dBm.
As can be seen from \autoref{fig:blinding_demonstration}, the output of the balanced detector clearly shows the effect of the blinding signal, with the modulated square wave being already visible at -15~dBm and becoming more and more pronounced as the power of the blinding signal is increased.
With a blinding power of -13~dBm the detector output jumps consecutively from -2048 to 2048 (the ADC used has 12-bit precision, this shows the signal jumping from the minimum to the max value that it can report).
This indicates that, in the case of our receiver, the blinded component is the ADC.
\par
\begin{figure}[h]
    \centering
	\begin{subfigure}{0.42\textwidth}
	   	\includegraphics[width=\linewidth]{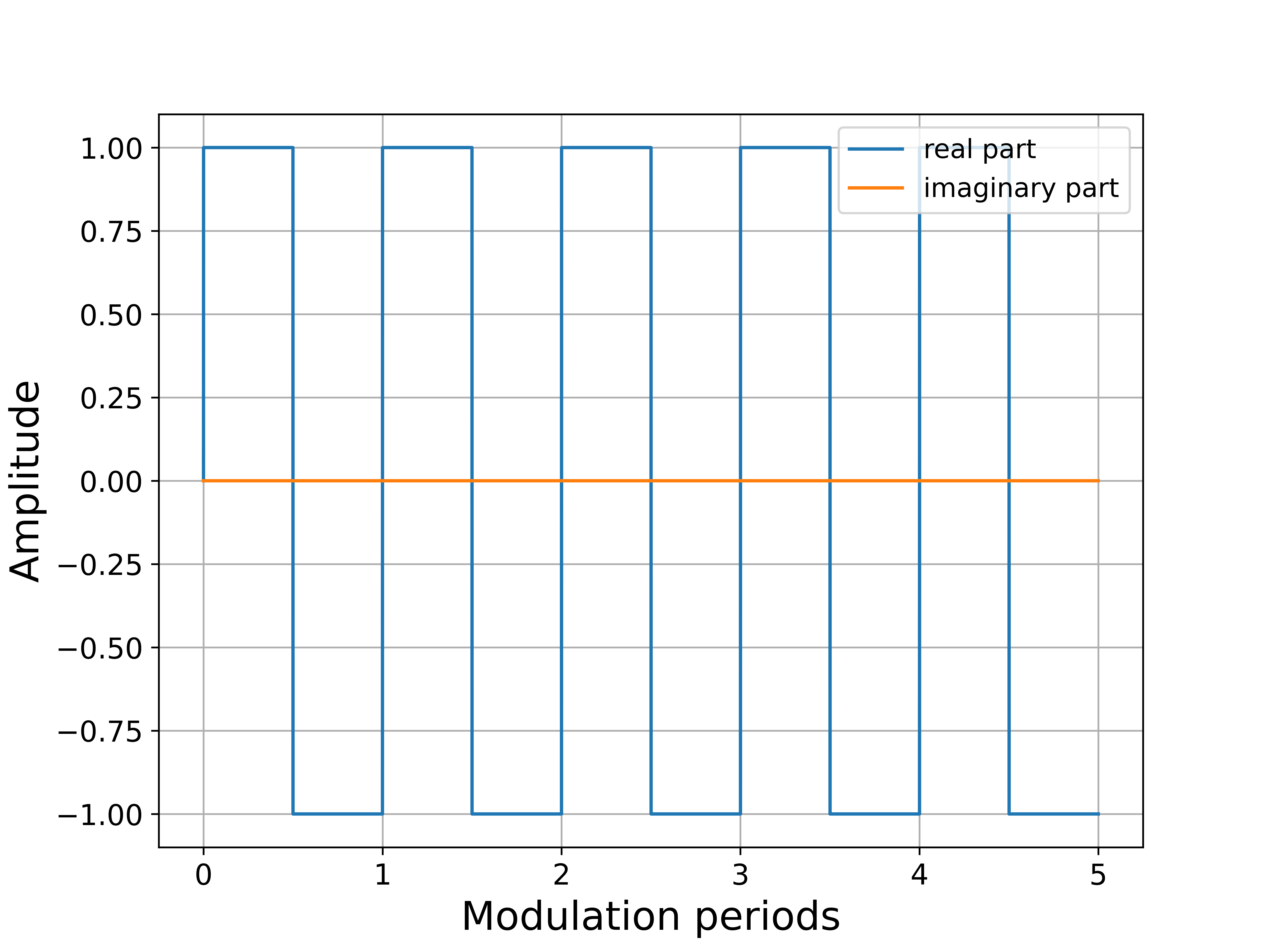}
   		\caption{Time domain representation of a square wave signal.}
	    \label{fig:square_wave_time}
   	\end{subfigure}
	\hspace{0.1\textwidth}
    \begin{subfigure}{0.42\textwidth}
	   	\includegraphics[width=\linewidth]{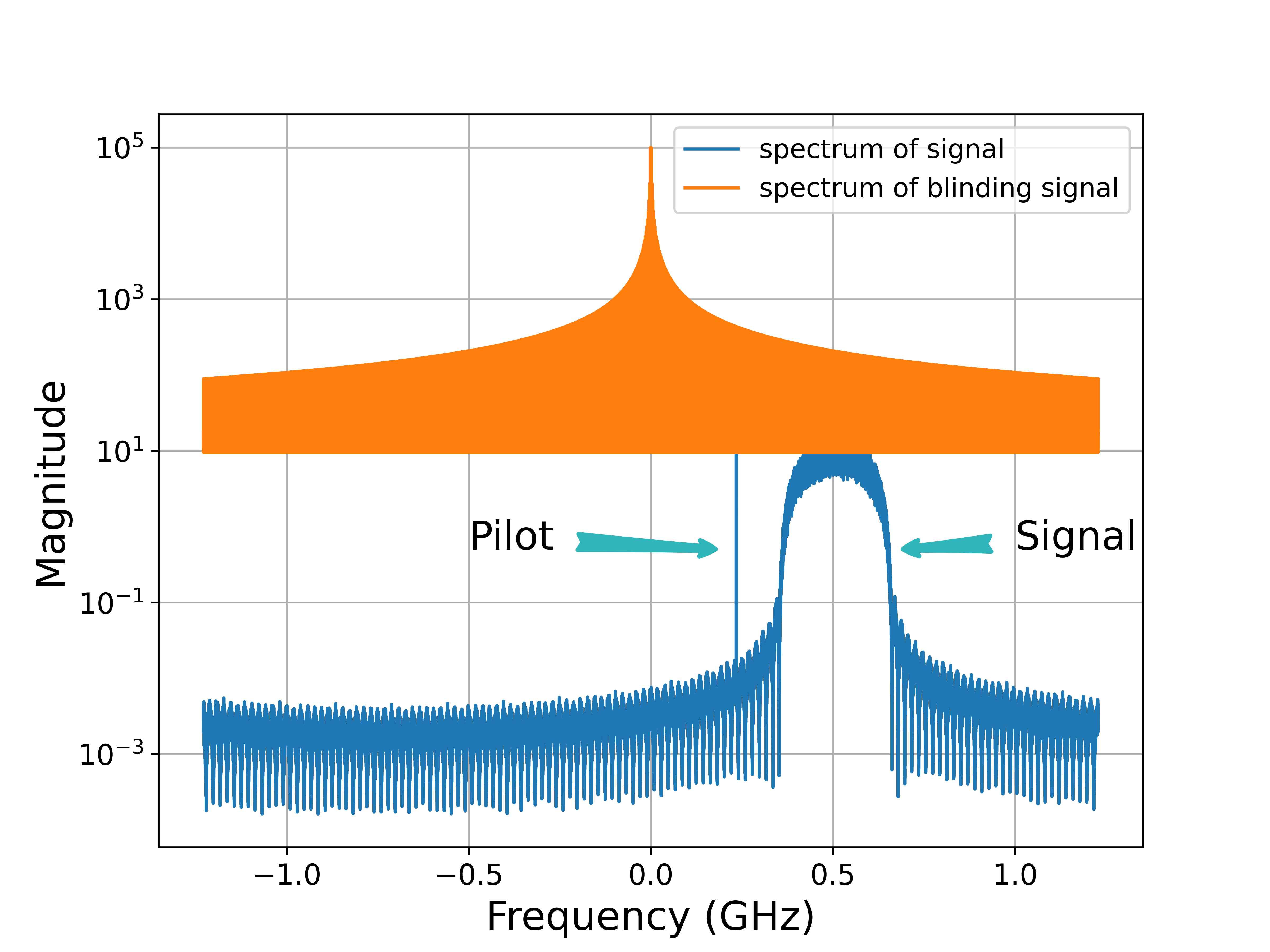}
   		\caption{Frequency domain representation of a square wave signal, superimposed on a simulated signal and pilot (indicated in plot).}
	    \label{fig:square_wave_spectrum}
   	\end{subfigure}
    \caption{Time domain and frequency domain representations of a square wave signal, used to illustrated the impact of a modulated blinding signal on the functioning of the CV-QKD system. Note that these graphs use simulated data generated to illustrate the concept.}
    \label{fig:square_wave}
\end{figure}

The modulation applied to the blinding signal will result in a widening of its spectrum, with possibly significant power being present at frequencies that are relevant for the functioning of the CV-QKD system.
This is illustrated in \autoref{fig:square_wave}, where a square wave signal is used as an example of a modulated blinding signal.
The square wave is a simple signal that can be easily generated by directly modulating the blinding laser, and it has a well known spectrum that contains many harmonics of the fundamental frequency, these harmonics can easily fall into the frequency range of interest for the CV-QKD system and impact its performance in a way that is not related to the saturation of the {Rx}.
This is seen in \autoref{fig:square_wave_spectrum}, where the spectrum of the square wave is shown to have significant power at the frequencies where the CV-QKD signal is located.
For the purposes of this illustration, the CV-QKD signal is simulated as a 64-QAM, RRC shaped signal, with a symbol rate of 153.6~MBaud and upconverted by 307.2~MHz, and the pilot is simulated as a tone at 38.4~MHz, the combined pilot and signal then being upconverted by a further 200~MHz, a value chosen purely for illustrative purposes.
\par
\begin{figure}[h]
    \centering
	\begin{subfigure}{0.42\textwidth}
	   	\includegraphics[width=\linewidth]{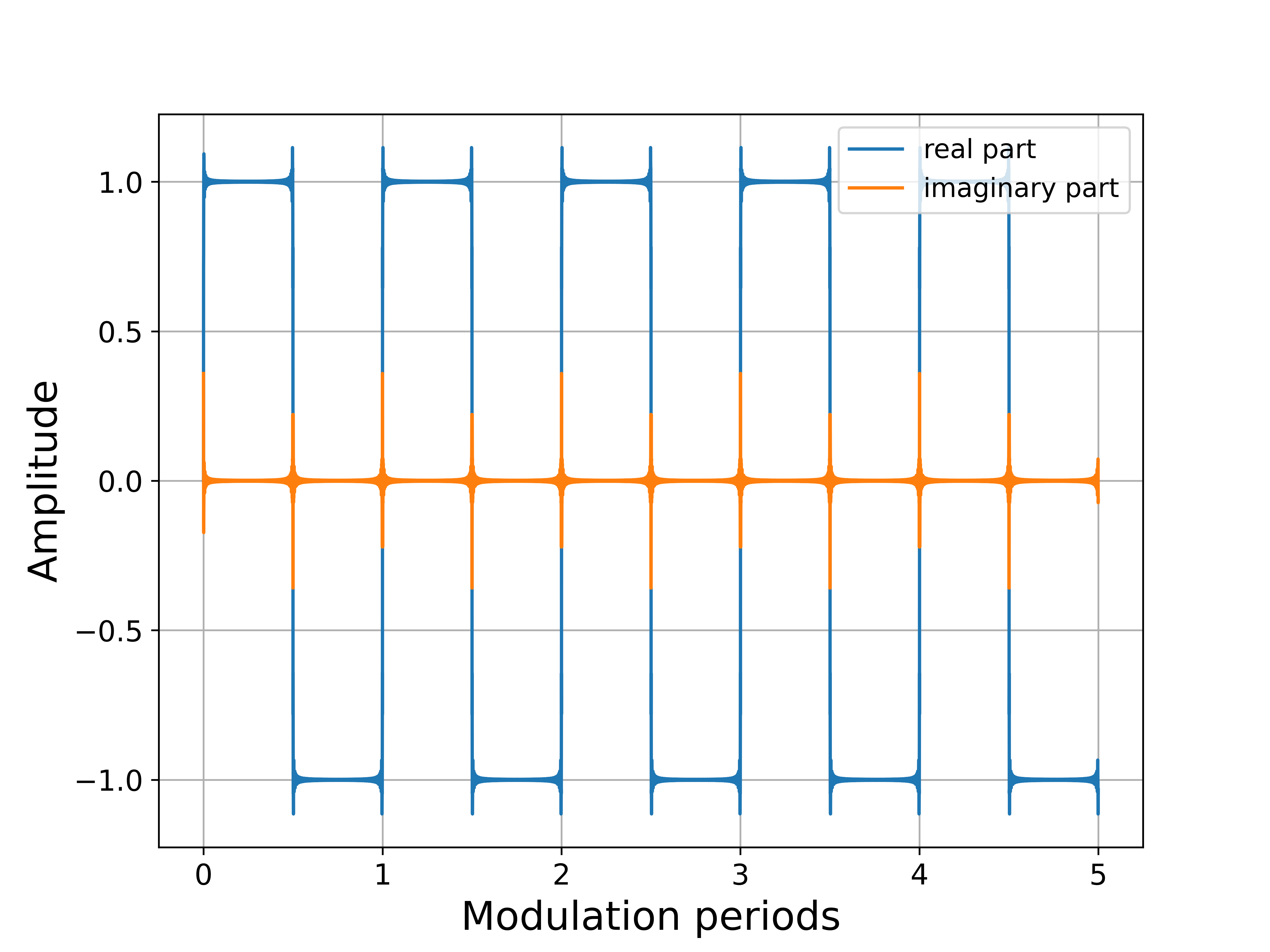}
   		\caption{Time domain representation of a square wave signal modulated as to avoid certain frequencies.}
	    \label{fig:square_wave_scooped}
   	\end{subfigure}
	\hspace{0.1\textwidth}
    \begin{subfigure}{0.42\textwidth}
	   	\includegraphics[width=\linewidth]{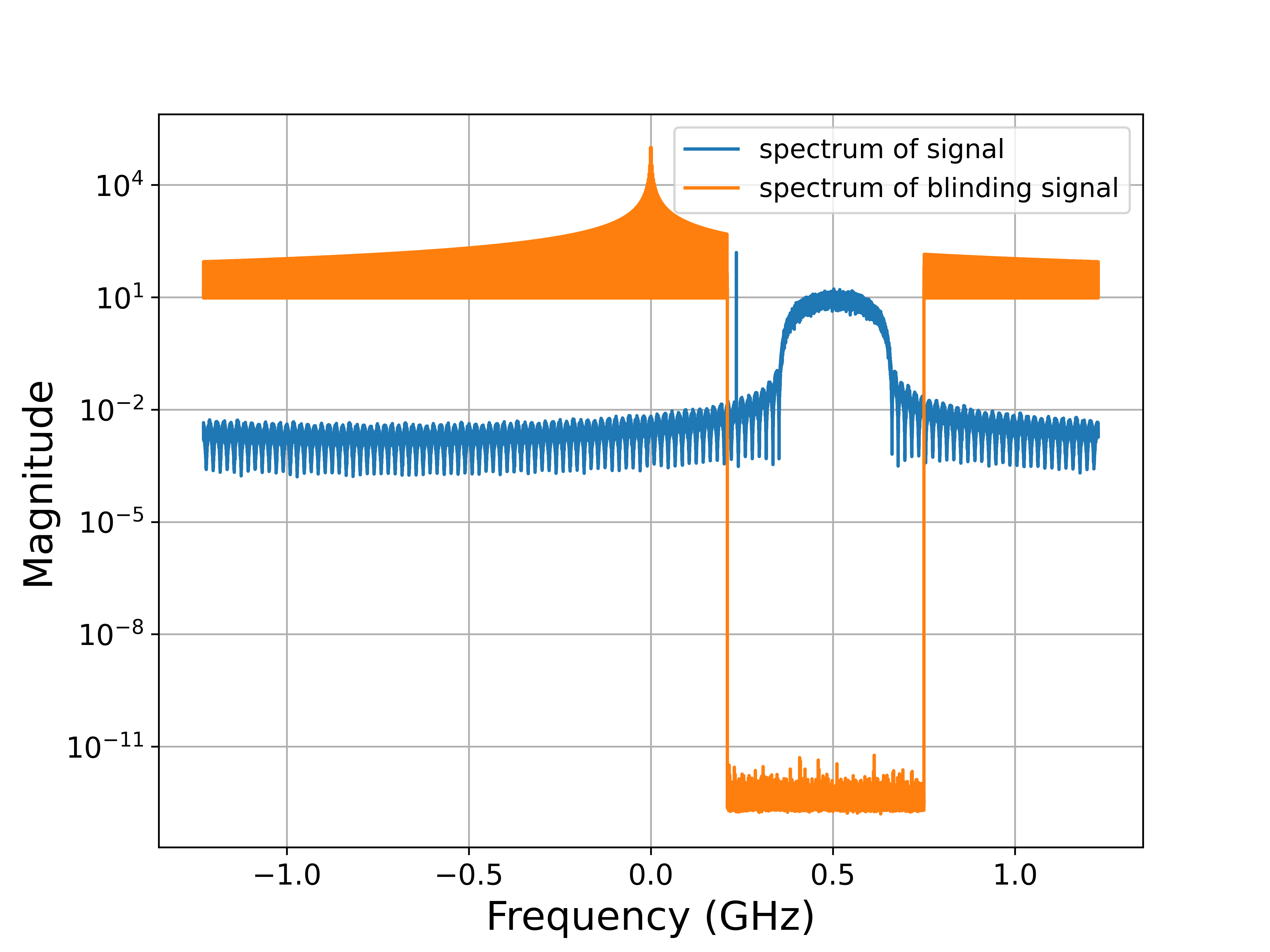}
   		\caption{Frequency domain representation of a square wave signal modulated as to avoid certain frequencies, superimposed on a simulated signal and pilot.}
	    \label{fig:square_wave_scooped_spectrum}
   	\end{subfigure}
    \caption{Time domain and frequency domain representations of a square wave signal modulated as to avoid certain frequencies, used to illustrate the impact of a modulated blinding signal on the functioning of the CV-QKD system. Note that these graphs use simulated data generated to illustrate the concept.}
    \label{fig:square_wave_scooping}
\end{figure}

Through careful tailoring of the modulation format, such as the one shown in \autoref{fig:square_wave_scooping}, this issue can be mitigated. 
To obtain this modulation pattern, the square wave in \autoref{fig:square_wave_time} was digitally filtered in order to remove the harmonics that fall into the frequency range of 210~MHz to 750~MHz, avoiding interference with the CV-QKD signal.
As is apparent from comparing \autoref{fig:square_wave_time} and \autoref{fig:square_wave_scooping},
this modulation can no longer be obtained by merely amplitude modulating the blinding signal, but requires also the usage of a phase modulator (or the use of an IQ modulator in place of the separate amplitude and phase modulators).
This shows that our method for bypassing the AC-coupling can be expanded to work multiple systems, provided that the modulation pattern is properly designed to avoid interference with the specific frequencies used by the CV-QKD system being targeted.
Furthermore, due to the combined frequency fluctuations of both lasers in {TX} and in the {RX} of the CV-QKD system, this modulation pattern will have to be continuously adapted in order to maintain the "notches" in the spectrum at the right frequencies.

\section{Security analysis and results}\label{sec:attack_results}
\begin{figure}[h]
    \centering
    \includegraphics[width=\linewidth]{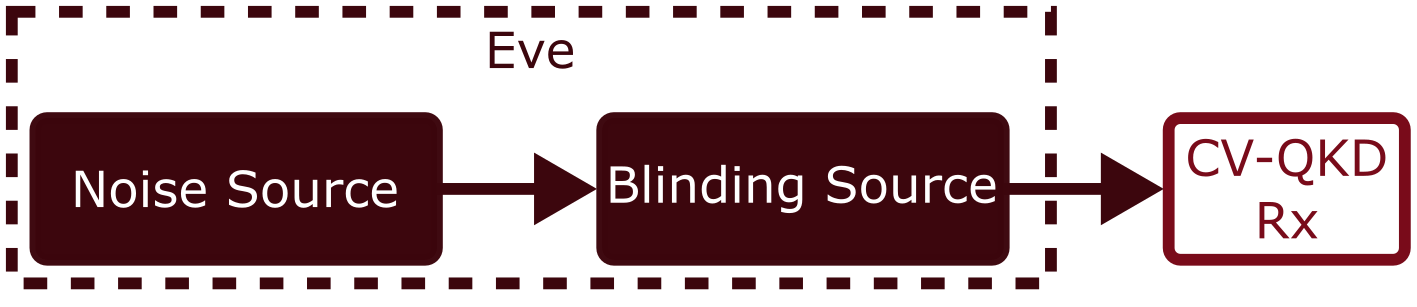}
    \caption{Diagram of experimental apparatus. In depth diagrams of the different components can be seen in \autoref{fig:DUT_diagram}, \autoref{fig:noise_source} and \autoref{fig:blinding_source}.}
    \label{fig:exp_diagram}
\end{figure}

To demonstrate the implemented attack, we assembled the apparatus shown in \autoref{fig:exp_diagram}.
This can be understood as a reduced version of the attack strategy shown in \autoref{fig:attack_strategy}, where only the receiver is present, and the exploitation stage is replaced by a noise source that simulates the impact of an attack on the channel parameters.
Internal details of each of the components can be seen in \autoref{sec:DUT}, \autoref{sec:noise_source} and \autoref{sec:blinding_source}.
\par
\begin{figure}[h]
    \centering
    \includegraphics[width=\linewidth]{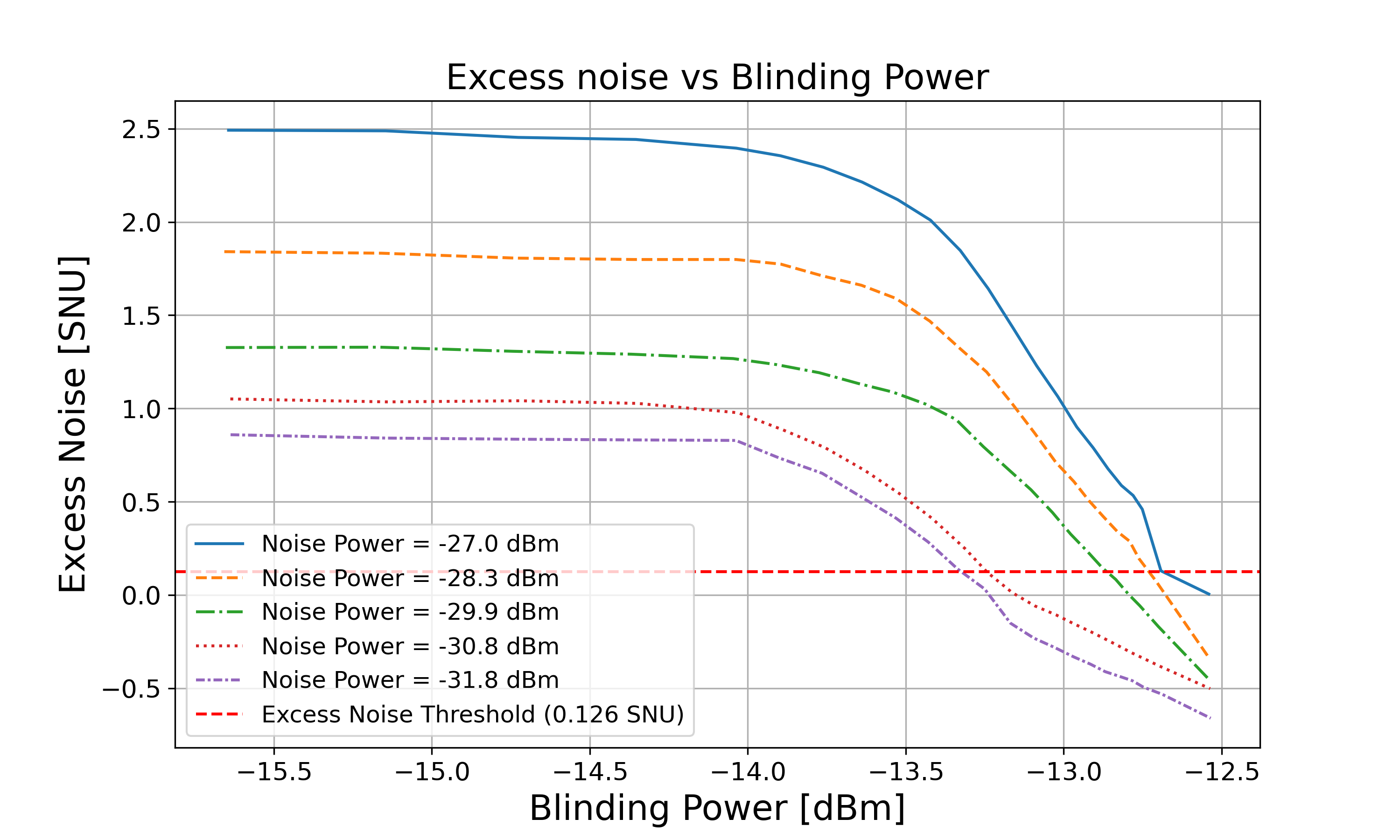}
    \caption{Impact of the blinding signal on the estimated excess noise.}
    \label{fig:attack_results}
\end{figure}

The power of the blinding signal was sweeped between -15.64536~dBm and -12.53858~dBm, while the noise source was set to introduce 5 different constant noise powers of -27.0~dBm, -28.3~dBm, -29.9~dBm, -30.8~dBm and -31.8~dBm (values already corrected to account for the 90/10 BS, matching the ones shown in \autoref{fig:noise_source_demonstration}).
The results of this experiment are shown in \autoref{fig:attack_results}, where the excess noise estimated by the receiver, following the method explained at the end of \autoref{sec:DUT}, is plotted against the power of the blinding signal for the different noise powers.
For low blinding powers, the excess noise estimated by the receiver closely follows the expected values from the preliminary characterization of the noise source, shown in \autoref{fig:noise_source_demonstration}, with the different noise powers being clearly distinguishable.
As the blinding power is increased up to -14~dBm, the excess noise curves shows little deviation, with the different noise powers still being clearly distinguishable.
One exception to this is the curve for noise power -27~dBm, which shows a slight decrease in the estimated excess noise, an effect that can be attributed to unequal saturation levels between the two arms of the receiver, resulting from a suboptimal polarization of the blinding signal.
After -14~dBm, the estimated excess noise starts to decline significantly, indicating that the detectors are becoming progressively more blinded and are no longer capable of correctly estimating the channel parameters.
This decline continues until the maximum blinding power is reached, with all but the curve for noise power -27~dBm showing a negative estimated excess noise, which indicates that the blinding is so high that even the receiver's thermal and shot noise are being suppressed.
\par
Also included in \autoref{fig:attack_results} is the maximum excess noise that can be tolerated by a CV-QKD system under the channel conditions assumed in the experiment (implementing a Probabilistic Shaped 64-QAM modulation format, using a Maxwell-Boltzmann distribution for the constellation points, at a channel transmission of 0.5, equivalent to 15~km of standard single mode fibre), which is shown as a dashed red line.
This value was computed using the security proof described in \cite{denys_explicit_2021} and corresponds to an excess noise of 0.126~SNU.
If the CV-QKD system estimates an excess noise value below this, it will result in a positive key rate being reported.
Given that the real key rate (i.e. the one measured without the blinding attack) would still be zero, the attack will have successfully extracted key information without being detected.
For our receiver, to successfully hide excess noises of 0.86~SNU, 1.05~SNU, 1.33~SNU, 1.84~SNU, and 2.49~SNU, blinding powers of -13.33~dBm, -13.24~dBm, -12.86~dBm, -12.73~dBm and -12.69~dBm, respectively, are sufficient.

\section{Conclusion}\label{sec:conclusion}

In this paper we have presented a novel implementation of coherent detector blinding attack against a complete CV-QKD receiver.
In our work we identified that the usage of an AC-coupled balanced detector effectively defeats the naive implementation of the blinding attack (using a constant blinding signal), as presented in~\cite{qin_homodyne-detector-blinding_2018,Qin2016}, but our results show that that is not sufficient to prevent a more sophisticated version.
Our experimental blinding source is able to achieve saturation even against receivers with AC-coupling, and can be tailored to work against different CV-QKD systems by adjusting the modulation pattern of the blinding signal.
Using our noise source, we simulated the impact of a strong collective attack on the channel parameters, introducing up to $\approx2.5$~SNU of excess noise.
We then showed that the excess noise introduced by this simulated attack can be hidden by the blinding source, thus demonstrating the feasibility of the enabling attack.
In assembling this blinding attack we have also identified novel countermeasures that do not require the usage of any new components.
Careful monitoring of the output of the receiver can be used to detect the presence of a blinding signal, for example by monitoring frequency components outside the bandwidth of the CV-QKD signal (effectively defeating the advanced modulation pattern shown in \autoref{fig:square_wave_scooping}), or by monitoring absolute values of the output signal and checking if these coincide with the saturation values of one of the receiver stages.
Note that the usage of optical filters and watchdog detectors remain as valid options, provided their optical bandwidth is wide enough to cover the relevant frequency range.

\section*{Acknowledgements}

The authors would like to thank Berhard Schrenk for his help in the assembly of the blinding source.

\printbibliography

\makeatletter
\let\firstofone\@firstofone
\makeatother

\end{document}